\documentclass[11pt,twoside]{article}

\usepackage{asp2006}
\usepackage{epsf}
\usepackage{graphicx}
\usepackage{lscape}

\begin{document}
\title{Star Formation Histories and Stellar Mass Growth out to $z> 1$}   %%% Fill in title
\author{Kai G. Noeske }  %%% Fill in author names
\affil{W.M. Keck Foundation Postdoctoral Fellow}
\affil{Harvard-Smithsonian Center for Astrophysics}    %%% Fill in author affiliations

\begin{abstract} %%% Abstract to run on from here.
The deepest multi-wavelength surveys now provide measurements of star
formation in galaxies out to $z>1$, and allow to reconstruct its history
for large parts of the galaxy population.  I review recent studies,
which have consistently revealed a picture where galaxy star formation
rates and their evolution are primarily determined by galaxy mass.
Unless they undergo a quenching of their star formation, galaxies of
similar masses have very similar star formation histories, which turn out
to be relatively smooth: star formation rates decline with redshift in a
primarily gradual manner, while typical starburst episodes have only
a modest amplitude that barely evolves.

I discuss how the found relations and their redshift evolution can provide
an observed reference star formation history as a function of galaxy
mass. 

The observed amplitudes and timescales of galaxy star formation are
not fully reproduced by current theoretical models, and are a
promising testbed to improve the assumed baryon physics. However,
measurements of star formation rates in distant galaxies need to be
treated with caution. Near-future data, methods and instruments will
help us to improve on calibrations and sensitivities for high
redshift star formation.
\end{abstract}

%%% MAIN BODY OF TEXT GOES HERE. CONSULT "INSTRUCTIONS FOR AUTHORS USING
%%% LATEX2E MARKUP", SECTIONS 2.3-2.6 FOR HELP WITH EQUATIONS, FIGURES,
%%% AND TABLES.

%\section{}   %%% Top level section head (remove "%" symbol)
%\subsection{}   %%% Second level section head (remove "%" symbol)
%\subsubsection{}   %%% Lowest level section head (remove "%" symbol)
%\section*{}    %%% Unnumbered top level section head (remove "%" symbol)
%\subsection*{}   %%% Unnumbered second level section head (remove "%" symbol)

\section{Star Formation and the Deep Multi-Wavelength Surveys}
\label{intro}
\vspace*{-1ex}

Star formation (SF) is responsible for most of those galaxy properties
that we can currently measure out to high $z$: luminosities, spectral
energy distributions, morphologies. Understanding the history and
physics of SF is fundamental for the understanding of baryons in
galaxies, and also for many other fields of astrophysics: the cosmic
evolution of gas and metals, the extragalactic background light, and
cosmological tests that rely on galaxies' clustering and number
densities to illuminate the evolution of Dark Matter structure.

Studies of SF histories have been dramatically advanced by the recent
arrival of deep, multi-wavelength surveys like GOODS, AEGIS and
COSMOS.  Their sensitivity allows to observe all galaxies down to
masses below typical $L^{\star}$ systems out to $z>1$, providing a
comprehensive picture of their evolution. Their variety of
multi-wavelength data, especially the Spitzer IRAC and MIPS $24\mu m$
data, have much improved the measurements of SF rates (SFR) and
stellar masses ($M_{\star}$) in distant galaxies, where dust
extinction corrections are challenging to measure
(e.g. \citet{daddi07}).

In the following, I summarize the first broad-brushed, but
comprehensive and new picture of SF in field galaxies that the deep
surveys have just revealed: SF was predominantly not driven by an
evolution of strong starbursts, but gradually declining on
mass-dependent scales
\citep{noeske07a,noeske07b,elbaz07,daddi07}. This picture ties
together some separate key results of the preceding decade, that (i)
the comoving SF rate (SFR) density of the Universe has decreased by
about an order of magnitude since $z=1$ \citep{madau96,hopkins04},
(ii) that many distant galaxies had high SFR that are unusual today,
and (iii) that the average SF history of galaxies is a strong function
of their mass \citep{cowie96,heavens04,juneau05}, a phenomenon dubbed
``Downsizing'' \citep{cowie96}.

Most of the following discussion is based on the AEGIS survey
\citep{davis07}. For more details, see \citep{noeske07a,noeske07b}.

\section{A Star Formation Rate-Stellar Mass Relation (``Galaxy Main
  Sequence'') out to $z\sim 2$}
\label{ms}
\vspace*{-1ex}

\begin{figure}[!t]
\centerline{\includegraphics[width=\textwidth]{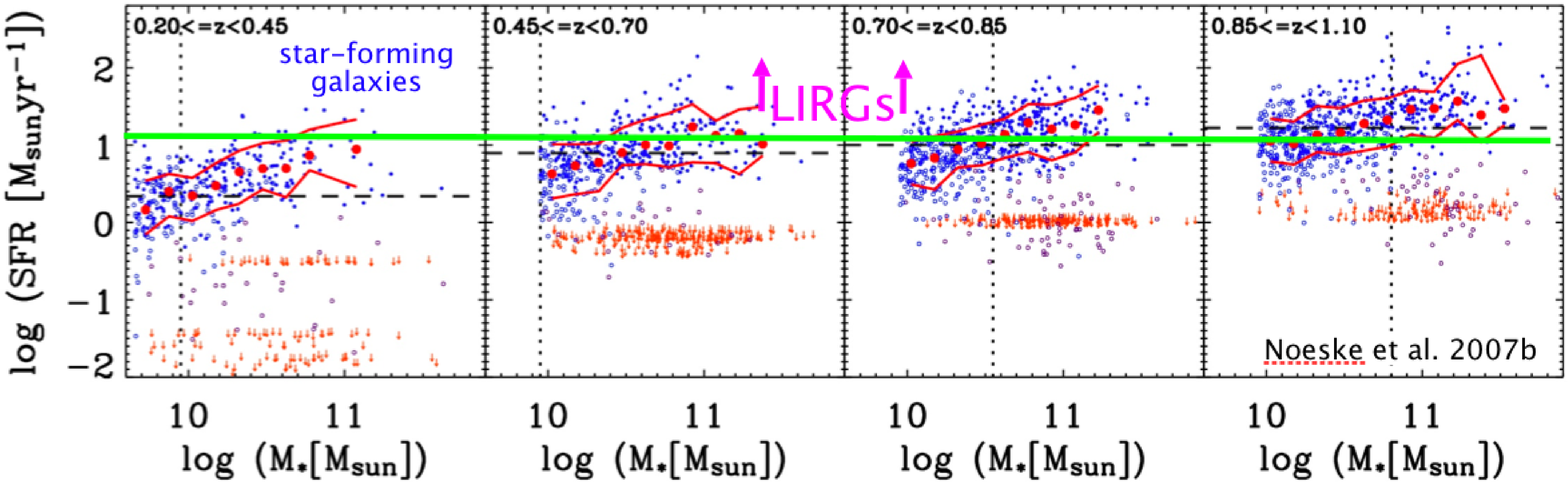}}
\vspace*{2ex}
\parbox{\textwidth}{\small Figure 1  Log(SFR) vs log($M_{\star}$) for
  $2905$ galaxies in AEGIS, in the $M_{\star}$ range where the data
  are $>80\%$ complete. See \citet{noeske07a} for details. The {\it
    dotted vertical line} marks $>95\%$ completeness. {\it Filled blue
    circles:} Combined SFR from MIPS 24$\mu$m and DEEP2 emission
  lines. {\it Open blue circles:} No 24$\mu$m detection, blue $U-B$
  colors, SFR from extinction-corrected emission lines. {\it Purple
    circles:} as open blue circles, but red $U-B$ colors, mostly
  LINER/AGN candidates. {\it Orange down arrows:} No robust detection
  of $f(24\mu$m$)$ or emission lines; conservative SFR upper limits
  shown. There is a distinct ``Main Sequence'' formed by fiducial SF
  galaxies (open and filled blue circles); galaxies with little or no
  SF lie below this sequence. Red circles: median of log(SFR) in mass
  bins of 0.15 dex for Main Sequence galaxies (blue circles). Red
  lines include 34\% of Main Sequence galaxies above and 34\% below
  the median of log(SFR), $\pm 1\sigma$ in the case of a normal
  distribution. {\it Horizontal black dashed line:} SFR corresponding
  to the $24\mu$m 80\% completeness limit at the center of each $z$
  bin. $24\mu$m-detected galaxies above the {\it green line} are
  LIRGs.}
\label{fig1}
\end{figure}

The evolution of SFR as a function of $M_{\star}$ and $z$ is
summarized in Figure 1, adapted from \citet{noeske07a}. Shown
are data from the AEGIS survey from $z=0.2$ to $1.1$. SFR are derived
from Spitzer $24\mu m$ photometry and DEEP2 emission lines
\citep{weiner07}, $M_{\star}$ from optical DEEP2 and NIR photometry
\citep{bundy06}. For other SFR tracers and calibrations, the results
are consistent, with small quantitative systematic differences.

The star-forming galaxies (blue symbols, predominantly late type
morphologies, mostly blue $(U-B)$ colors) segregate from those with no
measurable SF (red symbols; early types; red $(U-B)$ colors) and the
galaxies with weak emission lines that  are likely to have some residual SF
\citep{schiminovich07} or to be LINER/AGN-powered (purple; mostly early
types; red $(U-B)$ colors). See the caption of Figure 1 or
\citet{noeske07a}.

Importantly, the SF galaxies form a defined relation between SFR and
$M_{\star}$ over the whole $z$ range, with a spread in SFR at a given
$M_{\star}$ and $z$ that is crudely log-normal with a $1\sigma$ width
of $\la 0.3$\,dex (after correction for minimal estimates of SFR
errors) at all $z$.  Such a relation had been known at $z\sim 0.1$
\citep{brinchmann04}, and its existence to $z>1$ \citep{noeske07a} was
confirmed by \citet{elbaz07}. Recently, \citet{daddi07} reported this
relation with a similar spread in SFR at $z\sim 2$ (cf. also
e.g. \citet{foerster09}). Detailed studies at $z\sim 0.1$ are given in
\citet{schiminovich07} and \citet{salim07}.  While the scatter in SF
remains roughly constant at all observed $z$, the above authors find
the SFR at a given $M_{\star}$ to decrease by a factor of $\sim 6(20)$
from $z=1(2)$ to $0$ \citep{noeske07a,elbaz07,daddi07}.

For reasons explained in Section \ref{hrd}, we nicknamed the
SFR-$M_{\star}$ relation the ``Galaxy Main Sequence (GMS)''.

\section{Implications of the Galaxy Main Sequence: A New Picture of
  Star Formation in Field Galaxies since $z\sim2$.}
\vspace*{-1ex}

The surprising persistence of an equally sharp relation between SFR
and $M_{\star}$ out to $z\sim 2$, or over 10 Gyr in lookback time, has
profound implications for SF in field galaxies over most of the cosmic
time. These were first discussed in \citet{noeske07a} (and already in
part in \citet{zamojski07}), and pertain only to star-forming galaxies
on the SFR-$M_{\star}$ relation (and in the $z$-dependent $M_{\star}$
range where we are complete), not to those where SF was shut down by
still debated processes (cf. \citet{faber07}).

{\bf 1)} Galaxies of equal mass must have had similar SF histories,
else the scatter in SFR along the GMS would increase with time. The
smoothness of the dependence of SFR on $M_{\star}$ suggests that we
observe a generic mode of galaxy-wide star formation, possibly
dominated by the same set of few physical processes over several
decades in galaxy mass.

{\bf 2)} The $1\sigma$ spread of SFR at a given $M_{\star}$ and $z$ is
$\la \pm 0.3$ in log(SFR), and remains roughly equally narrow out to
$z\sim 2$. This finding limits the amplitude and duty cycles of
typical variations in SFR that galaxies can have experienced over the
past 10 Gyr: statistically, a galaxy spent 2/3 of its time within a
factor of 2 of its typical SFR at that $z$. If some galaxies underwent
stronger variations, causing much of the observed scatter, then the
remaining majority of galaxies must have had even smoother SF
histories. These limits on SFR variations constrain the effect of
galaxy interactions on galaxy SFRs; they are consistent with
theoretical predictions of the influence of frequent minor
interactions \citep{somerville01}, and constrain the longer-term
($10^8-10^9$\,yr) enhancement of SFR (e.g. \citet{cox06}) by major
mergers to a modest factor.

{\bf 3)} The factor by which the SFR along the GMS have decreased
since $\sim 2$ (see Section \ref{ms}) is much larger than the
amplitude of typical SFR variations. The {\em dominant} process in the
evolution of SF over the past 10 Gyr was hence {\em a gradual decline
  of SFR in individual galaxies}, with at most modest variations that
were superposed on that smooth decline.  It is especially noteworthy
that these SFR variations seem to have the {\em same relative
  amplitude} (factor) at all $z$: out to $z\sim 2$, episodic SF
variations or starbursts played a minor, barely evolving role in the
SF history of the Universe, and of typical galaxies. This result is
contrary to the formerly popular hypothesis that the evolution of SF
might be driven by increasingly frequent strong starbursts at higher
$z$.

The effect of galaxy interactions on SFR has now been measured for
galaxies in major and minor mergers, and close pairs (\citet{lin07};
\citet{robaina09}, and this conference; \citet{jogee09}, and this
conference). These studies have consistently shown a mild enhancement
of SFR: SFR distributions of interacting samples are shifted to $\la
2\times$ larger values than those of isolated control samples. This
limits the fraction of $M_{\star}$ formed at intermediate $z$ to $\la
10\%$ (\citet{robaina09}, and this conference).

Finally, Figure 1 (green line) reveals the origin of the
strong number density increase of Luminous Infrared Galaxies (LIRGs)
with $z$: apparently, galaxies become IR-luminous due to their
generic, gradual evolution of star formation, where SFR (and hence
likely their dust extinction) increase with $z$. This supports studies
\citep{bell05,melbourne05} that found LIRGs at intermediate $z$ to
have mostly regular, disk-like morphologies and suggested that LIRGs
are a universal phase in the intrinsic, gradual evolution of many
galaxies.

\section{The Galaxy Main Sequence: The Stellar Main Sequence -
  Equivalent for Galaxies}
\label{hrd}
\vspace*{-1ex}

\begin{figure}[!t]
\centerline{\includegraphics[width=0.6\textwidth,clip=]{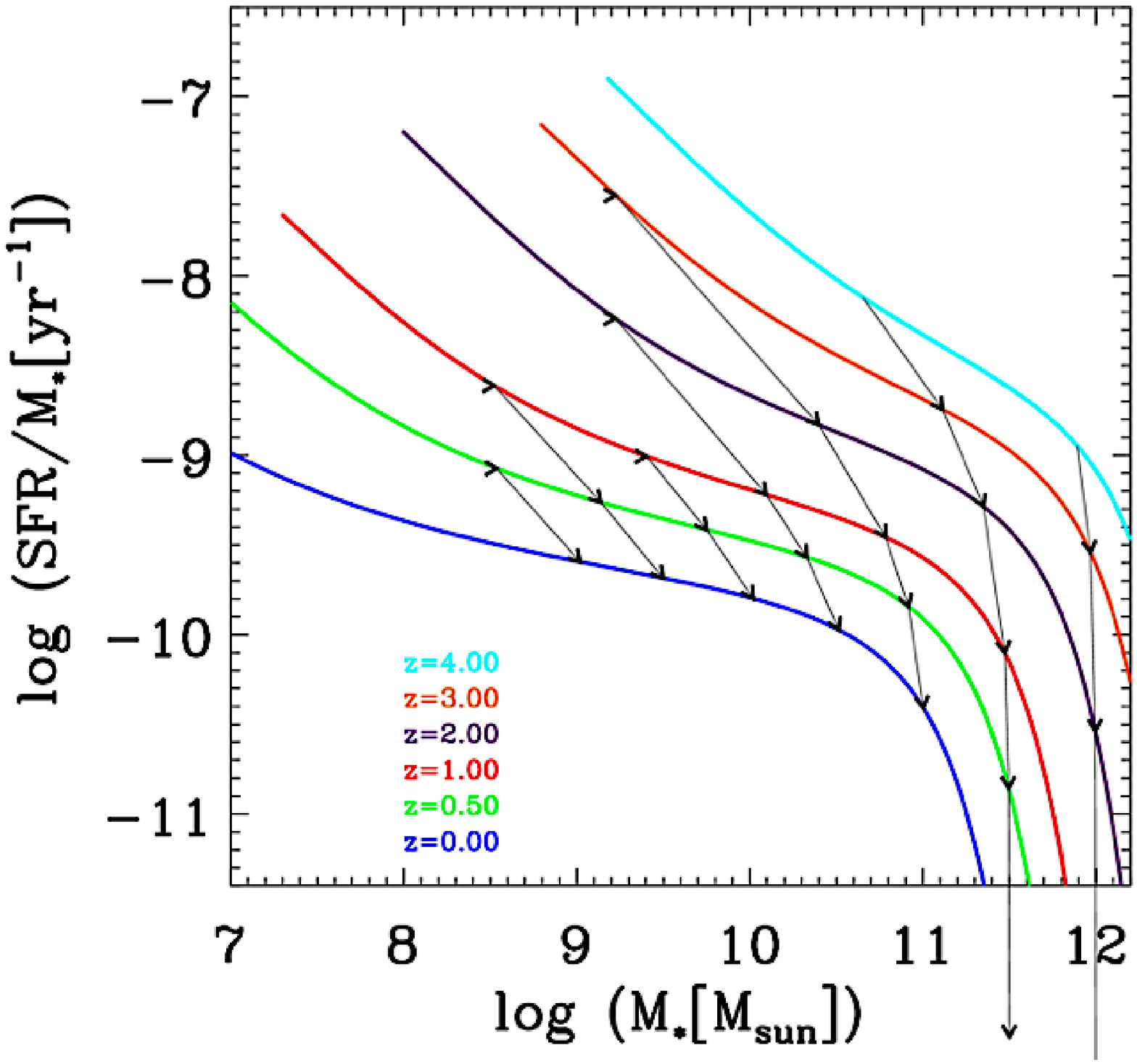}
\raisebox{0.3\textwidth}{\parbox{0.4\textwidth}{\small Figure 2 Isochrones in the
    log(specific SFR) - log(stellar mass) plane for the mass-dependent
    $\tau$ models presented in \citet{noeske07b}. Isochrones range
    from $z=0$ to $4$ from bottom to top; the arrows are evolutionary
    tracks for galaxies of different masses. Note the substantial
    stellar mass growth due to SF over time: galaxy properties cannot
    simply be compared for galaxies of equal stellar mass at different
    $z$, but require corrections for this mass growth \citetext{Noeske
      et al. 2009, in preparation}.}}
}
\label{fig2}
\end{figure}

The results I summarized in Section \ref{ms} reveal a fundamental role
of the SFR-$M_{\star}$ relation: because galaxies of equal mass have
similar SF histories, they must evolve along similar tracks in the SFR
- $M_{\star}$ plane. The SFR - $M_{\star}$ relation at a given $z$
must therefore mark the point on each mass-dependent evolutionary
track across the galaxy mass spectrum at that $z$: it is an {\em
  isochrone} cutting across the evolutionary tracks at a given
time. See Figure 2, where this is shown for the equivalent
case of specific SFR vs $M_{\star}$ at different $z$.

This is analog to another important isochrone in astrophysics - the
Hertz\-sprung-Russell Diagram, which is a superposition of the
mass-dependent stellar evolutionary tracks. In this picture, the
galaxies' SFR-$M_{\star}$ relation is the analog of the stellar main
sequence, where regular, active evolution, driven by the same set of
physical processes in an undisturbed system, proceeds until a change in
physics moves the object to its red late stages and passive end
stadium (the ``red and dead'' galaxies).  Incidentally, the galaxies'
`''Main Sequence Turnoff'' occurs systematically earlier for more
massive galaxies (e.g. \citet{bundy06}), similar to stars.

These similarities led us to adopt the term ``Galaxy Main Sequence'',
and the GMS is as fundamental to the understanding of galaxy evolution
as the stellar MS to stellar physics: from the GMS at different $z$,
we can recover the mass-dependent SF histories of galaxies.  In
\citet{noeske07b}, we presented a first simple, parametric approach:
The evolution of the GMS was modeled by simple, smooth model SF
histories, justified by the dominance of the smoothly declining
component of SF histories (Section \ref{ms}). We chose exponential SF
histories (``$\tau$ models''), given their previous success for many
applications; both parameters, the $e$-folding time $\tau$ and the
``formation redshift'' $z_f$ where SF begins, were allowed to depend
on the galaxies' ``baryonic mass'' as power laws \footnote{Note that
  the mass in \citet{noeske07b} is the ``baryonic mass'' of a closed
  box model. Since galaxies are not closed boxes, this mass will
  depend in a complicated way on the galaxies' actual
  (baryonic/dark/total/dynamical) masses. This ``mass'' merely acts as
  a dummy parameter to generate evolutionary tracks that correctly
  reproduce the data on SFR and $M_{\star}$ vs $z$, by keeping track
  of the stellar mass growth due to a SF history. These tracks can
  however easily be linked to actual stellar masses at any observable
  $z$, through the $M_{\star}$ they generate at a given $z$.}.  This
model reproduces the evolution of SFR and $M_{\star}$ on the GMS up to
$z=1.1$ remarkably well, and can attribute the scatter of SFR along
the GMS to scatter in SF history parameters at a given mass (see
\citet{noeske07b}, Figure 1), suggesting an even smaller role of
episodic or bursty SF. These $\tau$ models are the first
parametrization of the mass-dependent SF histories of galaxies.

Figure 2 shows the $\tau$ models (colored lines) in the
specific SFR (SFR normalized by $M_{\star}$) vs $M_{\star}$ plane. It
is equivalent to the SFR-$M_{\star}$ plane, essentially with the MS
rotated clockwise. For illustration, the model GMS is extrapolated out
to $z=4$; note that the models are only constrained by data to $z\sim
1$.  The evolutionary tracks (black arrows) reveal substantial mass
growth due to SF with redshift for all but the most massive galaxies,
also found by independent methods \citep{conroy07,zheng07}. Comparing
galaxies of equal $M_{\star}$ at different $z$ is therefore generally
not justified: one may compare very different objects. Instead, one
needs to compare galaxies on the same evolutionary track, i.e. apply a
mass correction that can for the first time be inferred from the
$\tau$ models discussed above, or future refined parametrizations.

\section{A Delayed Onset of Efficient Star Formation in Less Massive
  Galaxies}
\label{staging}
\vspace*{-1ex}

Interestingly, the $\tau$ model fits required both $\tau$ and $z_f$ to
be mass-dependent: Less massive galaxies had not only longer
$\tau$, i.e. a slower decline of SF, SF being less efficient and
having lower initial SFRs; they also had systematically later $z_f$,
equivalent to a later onset of SF.  The observed ``Downsizing'' in SF galaxies
is apparently a combination of both phenomena.

The late $z_f$ are required to account for the high specific SFR
(SFR/$M_{\star}$) of a majority of sub-$L_{\star}$ galaxies at $z\gg
0$ (see Figure 2 in \citet{noeske07b}. These imply ``doubling times''
much shorter than the age of the Universe, i.e. these galaxies cannot
have formed stars at their observed rate without overproducing their
stellar mass. The usual explanation for high specific SFR, starburst
events on top a lower SFR history, is not physical because a majority
of all such galaxies would need to simultaneously undergo a stochastic
event, and is also in contradiction with other observations - see
\cite{noeske07a} for details. In a substantial fraction - but not
necessarily all - of less massive galaxies, SF must hence have been
inefficient at early times and only attained sustained efficiency
later than in more massive galaxies (cf. Figure 3).

This mass-dependence of the onset of efficient SF, dubbed ``Staged
Galaxy Formation'' \citep{noeske07b}, is consistent with the observed
presence of very old stars --- roughly a Hubble time --- in the
majority of Local Group dwarf galaxies and other resolved systems. Our
data on SF to $z\sim 1$ only indicate that SF was inefficient, not
absent, in less massive galaxies, allowing for some old stars; (ii) efficient
SF only needs to delayed in the majority, but not all of such
galaxies; (iii) our data do not probe galaxies down to true dwarf
masses.

This systematic delay of efficient of SF in less massive galaxies is
not likely to be an artifact of SFR measurement errors. It is
consistent with statistical studies of galaxy SF histories from
independent methods - the evolution of stellar mass functions and the
fossil record in stellar populations of low $z$ galaxies;
cf. \citet{conroy07}, especially Figure 6; \citet{zheng07};
\citet{panter07}.

\begin{figure}[!t]
 \raisebox{0.\textwidth}{\parbox{0.65\textwidth}{\includegraphics[width=0.65\textwidth,clip=]{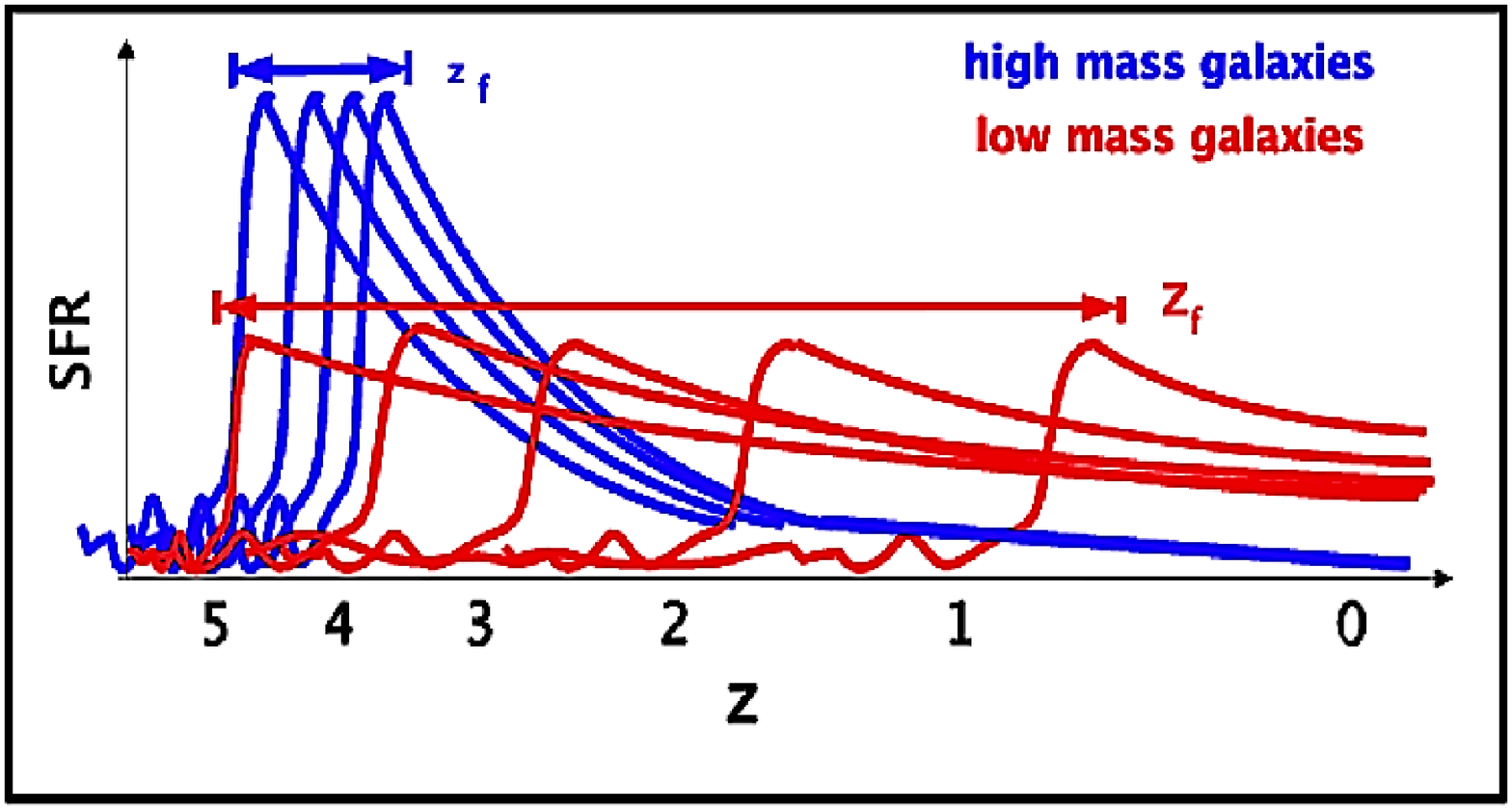}}}
  \raisebox{0.\textwidth}{\parbox{0.32\textwidth}{\small Figure
      3 Simplified cartoon sketch of the concept of {\it
        ``staged galaxy formation''} \citep{noeske07b}: In massive
      galaxies, SF declines on short timescales and begins to be
      efficient at high $z$. In less massive galaxies, SF declines not
      only more slowly; many low mass galaxies are also inefficient at
      forming stars at high $z$ and attain sustained efficient SF only
      at lower $z$.}}
\label{fig3}
\end{figure}

\section{Constraints to Galaxy Formation Models, and Uncertainties
  of SFR Measurements}
\vspace*{-1ex}

The various data on SFR and $M_{\star}$ out to $z>2$ have become an
important testbed for theoretical work on galaxy formation, and help
to improve the treatment of the complicated and numerically expensive
baryon physics. Current models of galaxy populations do generally
reproduce the SFR-$M_{\star}$ relation with a slope and scatter
similar to the observed one \citep{elbaz07,daddi07,dave08}.

Two types of discrepancies seem however to be universal between the
data and models: On the one hand, models underpredict the redshift
evolution of SFR for galaxies on the GMS. It is currently debated
whether this results from systematic errors in SFR measurements at
high $z$, problems of model physics, or both
\citep{elbaz07,daddi07,dave08}. On the other hand, the delay of
efficient SF in less massive galaxies is not correctly reproduced by
current models \citep{cirasuolo08,marchesini08,fontanot09}. Found from
independent data and methods (Section \ref{staging}), this difference
is probably physical and likely due to not yet fully understood
baryonic physics \citep{neistein06} that renders SF or its fueling
processes inefficient at early times in low mass halos.

While observed galaxy SFR across most of the cosmic time have provided
new key information for many purposes, considerable work is still
necessary (and underway!) to improve their calibrations and
systematics (see, e.g. \citet{salim09}), improve restframe IR coverage
with Herschel, ALMA and JWST. In addition, systematics like the
adopted stellar IMF and extinction curves can be tested from
non-standard derivations of SFR \citep{conroy07,chen09}.

{\small\acknowledgements %%% Text of acknowledgements runs on after this command.
This work received funding through grants from the W.M. Keck
Foundation, NASA and NSF and is based on observations with the
W.M. Keck Telescope, the Hubble Space Telescope, the Spitzer Space
Telescope, the Galaxy Evolution Explorer, the Canada France Hawaii
Telescope, and the Palomar Observatory.}

%%% THE BIBLIOGRAPHY
%%%
%%% CONSULT SECTION 3 OF "INSTRUCTIONS FOR AUTHORS" FOR HOW TO USE NATBIB.
%%% AUTHORS ARE ENCOURAGED TO USE EITHER THE "THEBIBLIOGRAPY" ENVIRONMENT
%%% BY UNCOMMENTING (DELETING THE "%" SYMBOL) THE COMMANDS BELOW, OR BY
%%% USING THE BIBTEX ENVIRONMENT. TO FIND OUT WHICH IS APPLICABLE TO YOUR
%%% CONTRIBUTION, CONSULT THE VOLUME EDITORS FOR YOUR PROCEEDINGS.
%%%

\end{document}